\documentclass[aps,prb,preprint]{revtex4}
\usepackage{amsmath}
\usepackage{amsfonts}
\usepackage{amssymb}
\usepackage{graphicx}

\begin{document}

\title{Noise-induced quantum coherence and persistent Rabi oscillations in a Josephson flux qubit}

\author{A.N. Omelyanchouk$^1$, S. Savel'ev$^{2,3}$, A.M. Zagoskin$^{2,3}$, E. Il'ichev$^4$,
Franco Nori$^{3,5}$}
\affiliation{$^1$B.Verkin Institute for Low Temperature Physics and Engineering,  61103, Kharkov, Ukraine\\
$^2$Department of Physics, Loughborough University, Leicestershire, LE11 3TU, United Kingdom\\
$^3$Digital Materials Laboratory, Advanced Science Institute,
RIKEN, Wako-shi, Saitama 351-0198, Japan\\ 
$^4$Institute of Photonic Technology (IPHT-Jena),
P.O. Box 100239, D-07702 Jena, Germany\\ $^5$Center for Theoretical Physics, Physics Department,
Center for the Study of Complex Systems,
The University of Michigan, Ann Arbor, MI 48109-1040, USA}

\begin{abstract}
We predict theoretically the enhancement of quantum coherence in a superconducting flux qubit by a classical external noise. 
First, the off-diagonal components of the qubit density matrix are increased. Second, in the presence of both ac drive and noise, the resulting Rabi oscillations survive ``in perpetuity", i.e. for times greatly exceeding  the Rabi decay time in a noiseless system. 
The coherence-enhancing effects of the classical noise can be considered as a manifestation of quantum stochastic resonance and are relevant to experimental techniques, such as Rabi spectroscopy.
\end{abstract}
\maketitle

\section{Introduction}

Rabi oscillations  are coherent periodic transitions between the states of a two-level quantum system with a low Rabi frequency $\Omega_R$, induced by a harmonic field in resonance with the much larger interlevel spacing $\Omega \gg \Omega_R$. They are among the most direct signatures of quantum behaviour.  Their observation in superconducting qubits was therefore a critical step in the direct proof of the qubits' behavior as controlled quantum objects and in the evaluation of their parameters (see, e.g., reviews \cite{Tsai-Nori,Zagoskin-Blais,Clarke-Wilhelm}). The decay rate of Rabi oscillations is determined by the relaxation and dephasing rates of the system and is quite fast, which makes their observation a nontrivial task, and thus subtracts from their usefulness as a qualitative criterion  of ``quantumness'' of a given device. 

This limitation is lifted when considering the {\em correlations} in a driven quantum system\cite{strong driving}. Then, as long as the coherence time in the system exceeds the Rabi period, the time correlations at Rabi period will reveal quantum coherence.  In particular, the spectral density of the qubit response will demonstrate a peak at the Rabi frequency $\Omega_R$. This {\em Rabi spectroscopy} was used to experimentally demonstrate the quantum behaviour in a driven flux qubit \cite{2} in its stationary regime. Alternatively, a simultaneous  driving of the qubit by a high-frequency ($\sim\Omega$) and a low-frequency ($\sim\Omega_R$) signal produces a resonant response when the frequency of the latter approaches $\Omega_R$ \cite{5}.

\subsection{Quantum correlations enhanced by classical noise}

In this paper we show that a classical noise, acting on the system, both reveals and enhances quantum correlations. By itself, classical noise   increases the off-diagonal elements of the qubit density matrix. If a regular ac drive is acting on the system in addition to the noise, we find  persistent Rabi oscillations: that is, a virtually non-decaying modulation of the fast, drive- and noise-induced oscillations of the density matrix elements, with a frequency close to the Rabi frequency.   
Both of these effects are related to   stochastic resonance, where 
noise reveals the instability of the system at some characteristic frequencies; thus, increasing the signal-to-noise ratio. There is a large literature on stochastic resonance, both in nonlinear classical systems \cite{1} and in quantum systems (see, e.g., \cite{Lofstedt,Grifoni,Wellens-Review} and references therein). In our case, the effect studied here can be understood  qualitatively as the noise occasionally ``resetting'' the Rabi oscillations, thus extending their lifetime to ``perpetuity'', as explained below. We also show that the details of the classical noise (e.g., using colored rather than white noise)  do not qualitatively affect the results.

In the calculations made for a superconducting flux qubit (e.g., \cite{Tsai-Nori,Zagoskin-Blais,Clarke-Wilhelm,2,5,6}) with  parameters consistent with   experimental data, we show that the quantum current fluctuations and quantum correlations in the qubit (related to the diagonal and off-diagonal components of its density matrix respectively) achieve a maximum at certain non-zero intensity of the classical external noise. 

\section{Model used}

A general quantum two-level system is described by the Hamiltonian (e.g., \cite{strong driving})
\begin{equation}
    \hat{H} = -\frac{\Delta}{2}\hat{\sigma}_x - \frac{\epsilon}{2}\hat{\sigma}_z \equiv
    \hat{H}_0 - \frac{\epsilon_1(t) +
    \delta\xi(t)}{2}\hat{\sigma}_z,
    \label{eq:1}
\end{equation}
where ${\sigma}_z$ and ${\sigma}_x$ are Pauli matrices, and the
eigenstates of ${\sigma}_z$ are the basis states in the localized
representation. The tunneling splitting energy 
$\Delta$ is usually determined by the geometry and fabrication details of the specific device, while the bias energy ${\epsilon}$ can be controlled externally and is split into three
components, $$\epsilon(t) = \epsilon_0 + \epsilon_1(t) + \delta\xi(t)$$
(static bias $\epsilon_0$, ac drive $\epsilon_1(t)$, and classical external noise $\delta\xi(t)$). In the
eigenbasis of $\hat{H}_0$ the Hamiltonian becomes
\begin{equation}
\hat{H}(t) = -\frac{\Omega}{2}\hat{\tau}_z -
\frac{1}{2}\left[\epsilon_1(t) + \delta\xi(t)\right]
\left(-\frac{\Delta}{\Omega}\hat{\tau}_x +
\frac{\epsilon_0}{\Omega}\hat{\tau}_z\right),
\label{eq:2}
\end{equation}
where $\hat{\tau}_{x,y,z}$ are Pauli matrices in the new basis,
and $$\Omega = \sqrt{\epsilon_0^2+\Delta^2}$$ is the static
interlevel distance.

\subsection{Noise effects}

Without loss of generality, we can assume that
all the external noise is produced by the variations of the
external magnetic flux in the qubit, i.e., $$\delta\xi(t) = \lambda\:
\delta\!f_n(t),$$ where $\lambda$ is a constant.
This noise can be thought of as produced by the control and readout circuitry.  
 We also take
$$\epsilon_1(t) = f_{ac}\sin \omega t.$$

The presence of the noise term in the Hamiltonian will naturally lead to a random noise source in the master equation for the density matrix. Its role is the same as of any explicitly time-dependent term describing the external field applied to the system. For any given realization of the random process $\delta\xi(t)$ the master equation and the density matrix are completely deterministic; the consequent averaging over the realizations is independent of quantum averaging.

We use the standard parametrization of the system's density matrix through the Pauli vector $(X,Y,Z)$:
 $$\hat{\rho} =
\frac{1}{2} (1+X\hat{\tau}_x+Y\hat{\tau}_y+Z\hat{\tau}_z),$$ to write the master equation,
$$\frac{d\hat{\rho}}{dt} =
-i[\hat{H}(t),\hat{\rho}]+\hat{\Gamma}\hat{\rho},$$ in the eigenbasis of the unperturbed Hamiltonian as 
\begin{eqnarray}
\frac{dX}{dt} = - CY - \Gamma_{\phi}X + \frac{\epsilon_0}{\Omega}Y \delta\xi(t); \nonumber\\
\frac{dY}{dt} = AZ + CX - \Gamma_{\phi}Y -
\left(\frac{\Delta}{\Omega}Z + \frac{\epsilon_0}{\Omega}X\right)
\delta\xi(t);\\
\frac{dZ}{dt} = -AY - \Gamma_r(Z-Z_{eq}) +
\frac{\Delta}{\Omega}Y \delta\xi(t) \nonumber.
    \label{eq:3}
\end{eqnarray}
Here we use the standard approximation for the dissipation
operator $\hat{\Gamma}$ in this basis. The dephasing and relaxation rates,
$\Gamma_{\phi}$ and $\Gamma_{r}$, characterize the intrinsic noise
in the system. Also, $$A = -\epsilon_1(t)\Delta/\Omega$$
and $$C = -\Omega - \epsilon_1(t)\epsilon_0/\Omega.$$ The quantity
$Z_{eq} = \tanh(\Omega/2T)$ is the equilibrium value of $Z$ at a
temperature $T$.

\section{Occupation probabilities}

The solutions of Eq.~(3) determine the occupation
probabilities of the upper (lower) level, $$P_±(t) = \frac{1}{2}(1\mp Z(t)),$$ and the quantum
coherence factors $X(t)$ and $Y (t)$. Any observable, 
 as well as its spectrum of fluctuations, can be
determined from $X(t)$, $Y (t)$ and $Z(t)$.  We will therefore
investigate the spectra of $X(t)$ and $Z(t)$, i.e., $S_X(\omega) =
X(\omega)$ and $S_Z(\omega) = Z(\omega)$.
 Under quite general assumptions, we can model the noise by  either a white (zero correlation time) or
a colored (finite
correlation time) Gaussian noise. 
For Gaussian white noise
 $$   \langle\delta\!f_n(t)\rangle = 0,\:\:    \langle\delta\!f_n(t) \delta\!f_n(t')\rangle = 2D\;\delta(t-t').$$
For colored noise with  correlation time $\tau$, the
stochastic process $\delta\!f_n(t)$ is determined by the equation \cite{7,8}
\begin{equation}
    \frac{d}{dt}\delta\!f_n(t) = -\frac{1}{\tau}\delta\!f_n(t) + \frac{1}{\tau}\zeta(t),
    \label{eq:6}
    \end{equation}
    where $\zeta(t)$ is a Gaussian white noise with $$\langle\zeta(t)\zeta(t')\rangle = 2D\zeta(t-t').$$
    
    \subsection{Dynamics with noise}
    
We solved numerically the system of Eqs.~(3) by the Ito method \cite{8}
for both white and colored external noise \cite{10}. For the numerical simulations, we used the parameters of a superconducting flux qubit 
(e.g., \cite{6}). 
 This choice was made because the flux qubit is a  well-understood and throughly investigated device. In particular, there exists a quantitative theory of its response to a low-frequency drive \cite{Greenberg02a,
3, 5}.   The experimental techniques necessary for the observation of the phenomena we consider here are also well developed  and yield  results which so far are in  excellent agreement with theory \cite{2}.

\section{Superconducting flux qubits}

A superconducting flux qubit consists of a superconducting loop
interrupted by three Josephson junctions \cite{Tsai-Nori,Zagoskin-Blais,Clarke-Wilhelm}. The state of
the qubit is controlled by the applied magnetic flux $\Phi_e =
f_e\Phi_0$ through the loop (where $\Phi_0 = h/2e$ is the flux
quantum). In the vicinity of $f_e = 1/2$, the ground state of the
system is a symmetric superposition of states $|L\rangle$ and
$|R\rangle$, with clock- and counterclock-wise circulating
superconducting currents of amplitude $I_p$, respectively. The qubit is described by the Hamiltonian (1) in the basis
$\{|L\rangle$; $|R\rangle\}$, while  in the eigenbasis it is described by the Hamiltonian (2). The bias
$$\epsilon = I_p\Phi_0(f_e-1/2)$$ is tunable, while the tunneling
amplitude $\Delta$ is determined by the fabrication of the loop
and the junctions. The circulating current,
\begin{equation}
    I(t) = I_p \left[\frac{\Delta}{\Omega}X(t) - \frac{\epsilon_0}{\Omega}Z(t)\right],
    \label{eq:4}
\end{equation}
is an observable and can be detected by the impedance measurement technique
\cite{4}. In this approach, a high-Q resonant tank circuit is
coupled to  the qubit. The tank's impedance can be measured with high precision and is influenced by the
qubit current  and/or its fluctuations. Therefore the  low-frequency changes in the qubit current can be directly measured (like in the so-called Rabi
spectroscopy \cite{2}).

We use the following parameters for the qubit: $I_p\Phi_0 = 200$
GHz, and $\Delta =$  1.4 GHz, which are consistent with typical
experiments \cite{Zagoskin-Blais,2,4}. We also assume reasonable, even somewhat pessimistic, values
for $\Gamma_r$ and $\Gamma_{\phi}$ (both equal to 0.1 GHz).  Note that for this choice of
decoherence rates there appear virtually no spectral features in
the absence of  external noise. Our results are presented in
Figs.~1 and 2. The data for the power spectrum are averaged over 50 random realizations of the random source in Eqs.~(\ref{eq:3}).

\subsection{Noise-enhanced, not dissipation-enhanced,  quantum coherence}

For white external noise and no ac drive (Fig.~1a) we see
that the spectrum of the coherent part of the qubit density
matrix, $S_X(\omega)$, exhibits a response reminiscent of classical stochastic resonance: as the
noise intensity $D$ {\it increases} from $10^{-7}$ to $10^{-4}$
GHz$^{-1}$, the maximum value of $S_X(\omega)$ goes through a
well-defined {\it maximum}. The noise color suppresses the peak
amplitude (at the same noise intensity), but does not shift its
position as a function of frequency from the characteristic frequency corresponding to the interlevel splitting (Fig.~1b). This  effect corresponds to {\it ``noise-enhanced quantum coherence''}. Unlike ``{\em dissipation}-enhanced quantum coherence'' (observed in, e.g., NMR experiments~\cite{Viola2000} and Bose-Einstein condensate experiments~\cite{Witthaut2009}) and stochastic resonance in spin chains (theoretically studied in~\cite{Huelga,Rivas}),  the noise we consider here is {\em not} due to intrinsic fluctuations in the system, and is therefore independent of the relaxation and dephasing rates.

A similar situation
arises in the presence of a periodic drive (Fig.~2a). Here, as the
noise intensity increases, the spectral density of $Z$   also grows
initially, and then decreases. Similarly, the colored noise
suppresses the peak amplitude, but does not change its
position (Fig.~2b) (the second, sharp peak in $S_Z(\omega)$ is due
to resonant interlevel transitions). Note that, in the absence of the
external noise, with the chosen
decay and dephasing rates $\Gamma = 0.1$,  Rabi oscillations quickly decay and do not
show on the spectrum. The noise forces oscillations with the Rabi
frequency, which produce a peak in the spectral density, thus
revealing  Rabi oscillations for very long times (similar to the experiment in Ref.~\cite{2}).
This phenomenon could be considered as another example of {\em quantum} stochastic resonance, as opposed to its classical counterpart.

\section{Conclusions}

Our numerical simulations show that quantum
stochastic resonance in a  qubit  manifests itself as a
resonant enhancement of the spectrum of the coherent part of the density matrix, induced by the
external classical noise. In the absence of an external ac drive, the noise enhances the off-diagonal matrix elements of the density matrix, while in a driven qubit it leads to very long-lived Rabi oscillations with a randomly shifting phase.
This effect is predicted to show up in the
Rabi spectroscopy of superconducting flux qubits and provides
another nontrivial signature of quantum coherence \cite{22,23}, which
can be observed in a stationary regime, notwithstanding its formally finite
characteristic decay time. The results of this work are quite general and apply to any quantum two-level system affected by an external classical noise (e.g., \cite{24,25}).
\\

{\bf Acknowledgements}
\\

We acknowledge partial support from the National Security Agency, LPS, Army Research Office, National Science Foundation,
JSPS-RFBR 06-02-91200, MEXT Grant-in-Aid No.~18740224, JSPS-CTC
program and FRSF (grant F28.21019). We are grateful to Fabio Marchesoni and Alexander Balanov  for many
valuable discussions. 

ANO and EI  appreciate the hospitality of DML, ASI, RIKEN. 
EI gratefully acknowledges financial support from the EC through
the EuroSQIP project, the Federal Agency on Science and Innovations of the Russian Federation under contract No 02.740.11.5067, and partial   support by the DFG IL150/6-1.
\\

{\bf References}
\\
\begin{figure}[btp]
\begin{center}
\includegraphics*[width=12.0cm]{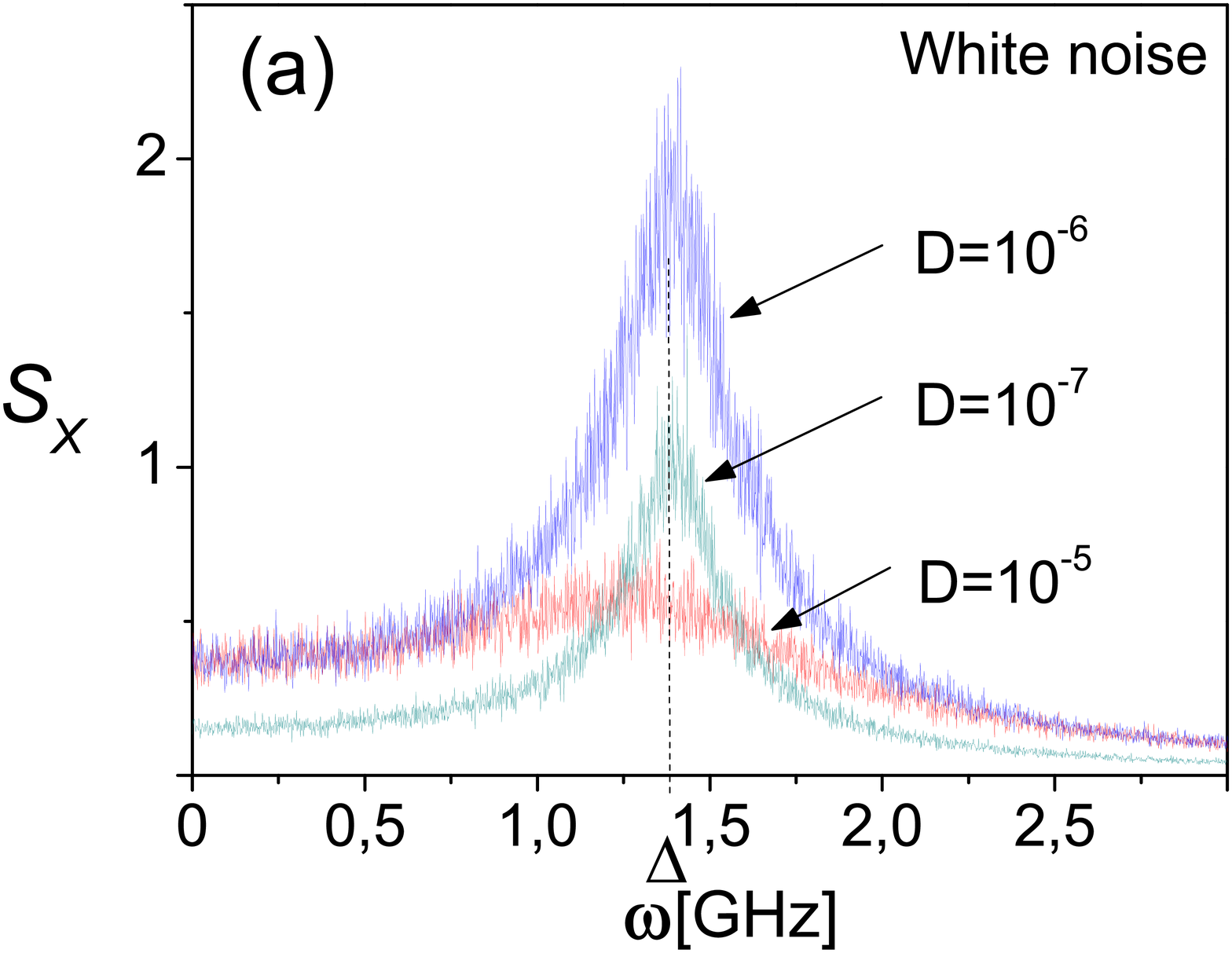}\\
\includegraphics*[width=12.0cm]{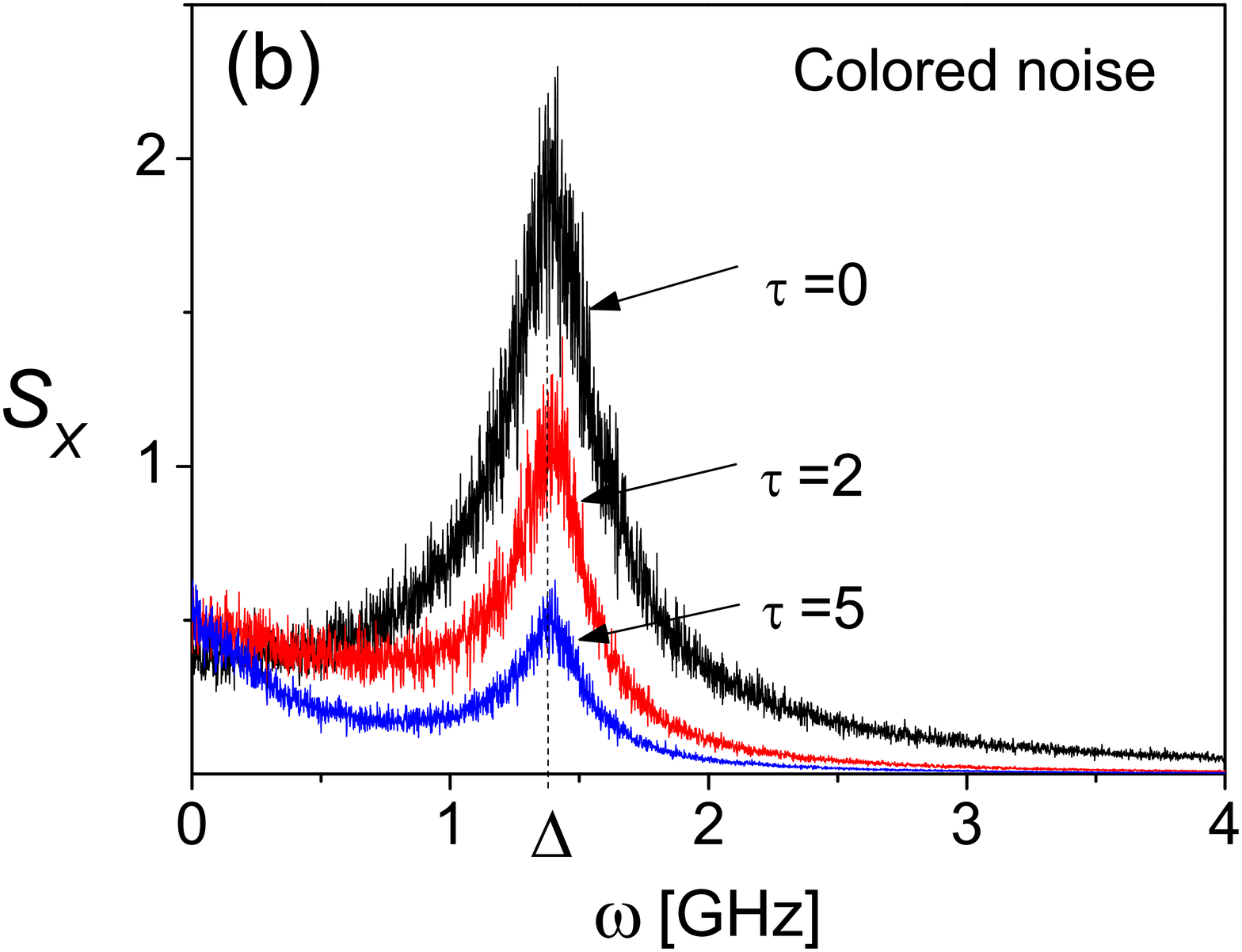}
\end{center}
\caption{(Color online) (a) Spectral density of $X$, $S_X(\omega)$, at the
optimal point with {\em no} ac signal ($f_{dc} = 0.5; \: f_{ac} = 0$) in
the presence of external white noise with intensity $D$ (in
GHz$^{-1}$). Note that noise can {\em enhance} the signal. (b) $S_X(\omega)$ under the same conditions, but in
the presence of a colored noise ($\tau = 0, 2, 5$) with
intensity $D = 10^{-6}$ GHz$^{-1}$. The vertical axes have been multiplied by 100.}
\end{figure}

\begin{figure}[btp]
\begin{center}
\includegraphics*[width=12.0cm]{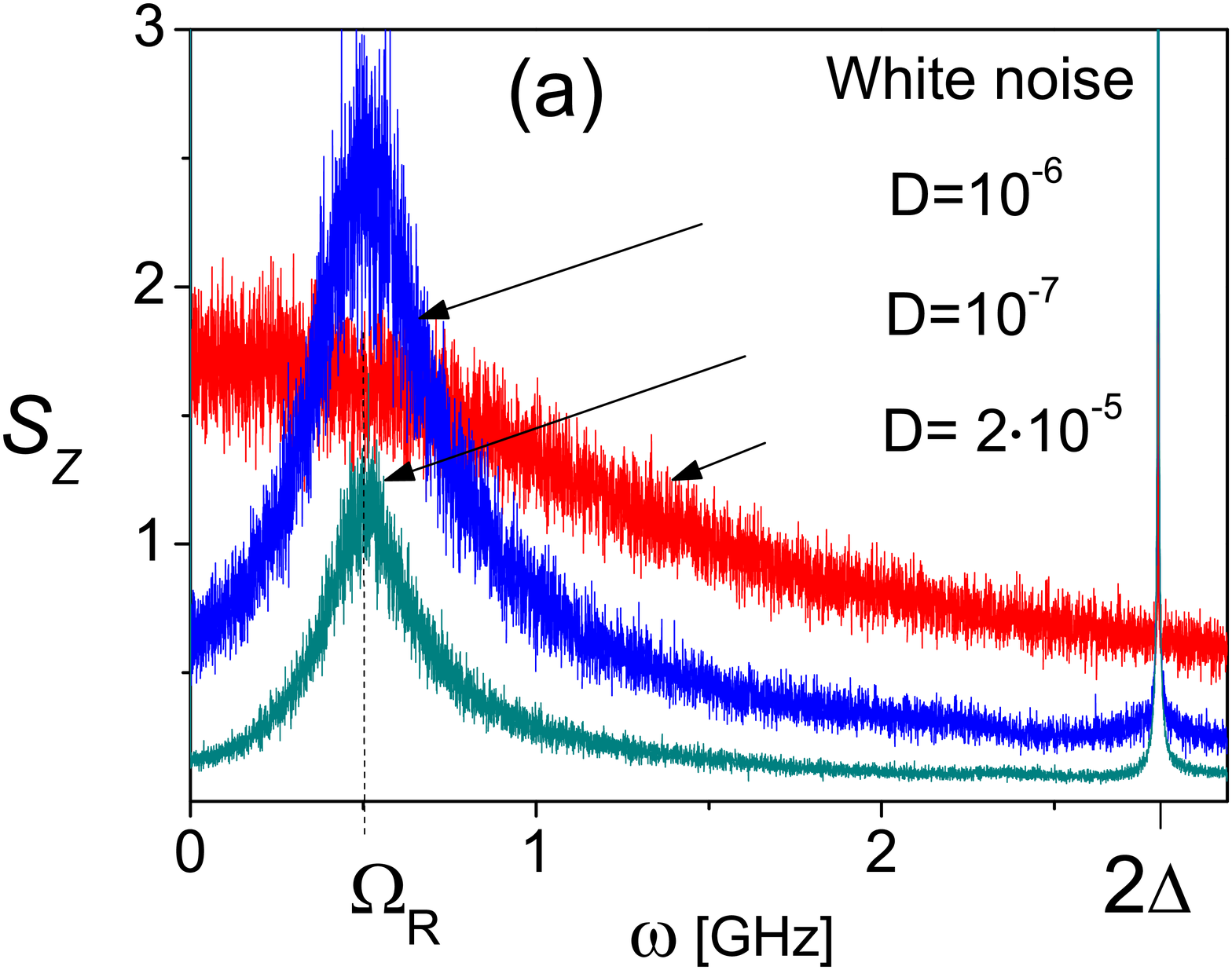}\\
\includegraphics*[width=12.0cm]{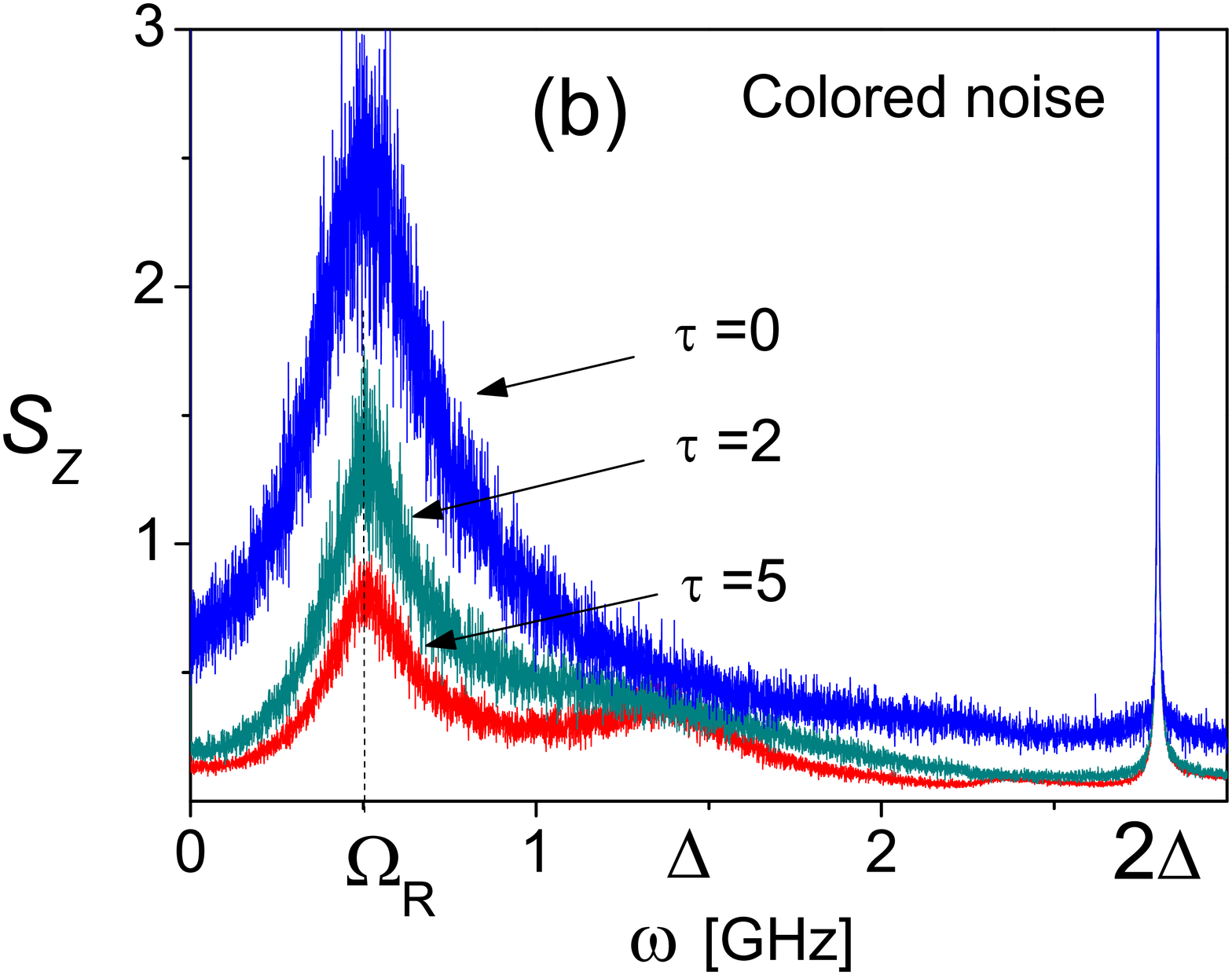}
\end{center}
\caption{(Color online) (a) Spectral density $S_Z(\omega)$ of the $z$-component of the Bloch vector  at the
optimal point $f_{dc} = 0.5$, {\em with} an applied ac signal $f_{ac} =
0.005$ in the presence of external white noise with intensity $D$
(in GHz$^{-1}$). Note that noise can actually enhance the signal. (b) $S_Z(\omega)$ under the same conditions, but
in the presence of colored noise ($\tau = 0, 2, 5$) with
intensity $D = 10^{-6}$ GHz$^{-1}$. The vertical axes have been multiplied by 1000.}
\end{figure}


\begin{thebibliography}{99}
\bibitem{Tsai-Nori} J.Q. You and F. Nori, Physics Today {\bf 58}, No.11, 42 (2005).
\bibitem{Zagoskin-Blais} A. Zagoskin and A. Blais, Physics in Canada, {\bf 63}, 215 (2007) 
\bibitem{Clarke-Wilhelm} J. Clarke and F.K. Wilhelm, Nature {\bf 453}, 1031 (2008).
\bibitem{strong driving} S. Ashhab, J.R. Johansson, A.M. Zagoskin, and F. Nori, Phys. Rev. A {\bf  75}, 063414 (2007).   
\bibitem{2}  E. Il'ichev, N. Oukhanski, A. Izmalkov, T. Wagner, M. Grajcar, H.-G. Meyer, A.Yu. Smirnov, A. Maassen van den Brink, M.H.S. Amin, and A.M.  Zagoskin, Phys. Rev. Lett. {\bf 91}, 097906 (2003).
\bibitem{5} Ya. S. Greenberg, E.Il'ichev, and A.Izmalkov, Europhysics Letters, {\bf 72}, 880 (2005). 
\bibitem{1} L. Gammaitoni, P. H\"{a}nggi, P. Jung, and F. Marchesoni, Rev. Mod. Phys. {\bf 70}, 223 (1998).
\bibitem{Lofstedt} R. Lofstedt and S.N. Coppersmith, Phys. Rev. Lett. {\bf 72}, 1947 (1994).
\bibitem{Grifoni} M. Grifoni and P. H\"{a}nggi, Phys. Rev. Lett. {\bf 76}, 1611 (1996).
\bibitem{Wellens-Review} T. Wellens, V. Shatokhin, and A. Buchleitner, Rep. Prog. Phys. {\bf 67}, 45 (2004).
\bibitem{6} J.E.  Mooij, T.P. Orlando, L. Levitov, L. Tian, C.H. van der Wal, and S. Lloyd, Science {\bf 285}, 1036 (1999).
\bibitem{7} P. H\"{a}nggi, P. Jung, C. Zerbe, and F. Moss, Journ. Stat. Phys. {\bf 70}, 25 (1993).
\bibitem{8} C.W. Gardiner, Handbook of stochastic methods, 2nd ed. (Springer, Berlin, 1990).
\bibitem{10} In these calculations the frequency is taken in GHz. The sample
time $\tau_{\rm sample} = 15000$, the number of steps $N=2^{20}$, and the sampling
frequency $f = N/\tau_{\rm sample}$.
\bibitem{Greenberg02a} Ya.S. Greenberg, A. Izmalkov, M. Grajcar, E. Il'ichev, W. Krech, H.-G. Meyer, M.H.S. Amin, and A. Maassen van den Brink, Phys. Rev. B \textbf{66}, 214525 (2002).
\bibitem{3} A.Yu. Smirnov, Phys. Rev. B {\bf 68}, 134514 (2003).
\bibitem{4} E. Il'ichev, N. Oukhanski, T. Wagner, H.-G. Meyer, A.Yu. Smirnov, M. Grajcar, A. Izmalkov, D. Born, W. Krech, and A. Zagoskin, Low Temp. Phys. {\bf 30}, 620 (2004).
\bibitem{Viola2000} L. Viola, E. M. Fortunato, S. Lloyd, C.-H. Tseng, and D. G. Cory, Phys. Rev. Lett. {\bf 84}, 5466 (2000).
\bibitem{Witthaut2009} D. Witthaut, F. Trimborn, and S. Wimberger, Phys. Rev. A {\bf 79}, 033621 (2009).
\bibitem{Huelga} S.F. Huelga and M. Plenio, Phys. Rev. Lett. {\bf 98},170601 (2007).
\bibitem{Rivas} A. Rivas, N.P. Oxtoby, and S.F. Huelga, Eur. Phys. J. {\bf 69}, 51 (2009).
\bibitem{22}  A.N. Omelyanchouk, S.N. Shevchenko, A.M. Zagoskin, E. Il'ichev, and F. Nori, Phys. Rev. B {\bf 78}, 054512 (2008).
\bibitem{23} S.N. Shevchenko, A.N. Omelyanchouk, A.M. Zagoskin, S. Savel'ev, and F. Nori, New J. Phys. {\bf 10}  073026 (2008).
doi:10.1088/1367-2630/10/7/073026
\bibitem{24} A.M. Zagoskin, S. Ashhab, J.R. Johansson, and F. Nori, 
Phys. Rev. Lett. {\bf 97}, 077001 (2006).
\bibitem{25} S. Ashhab, J.R. Johansson, and F. Nori,
New J. Phys. {\bf 8}, 103 (2006).
\end{thebibliography}
\end{document}